\documentclass{JINST}
\usepackage{subfig}
\usepackage{mathptmx}
\usepackage{helvet}
\usepackage{amsmath}
\usepackage{units}
\usepackage{url}
\usepackage[norounding,point]{rccol}
\usepackage{amssymb}
\usepackage[multiple]{footmisc}
\usepackage{threeparttable}
\newcommand{\ud}{\,\mathrm{d}}

\title{Correction of beam-beam effects in luminosity measurement in the forward region at CLIC}
\author{S. Luki\'c\thanks{Corresponding
author.}~, I. Bo\v{z}ovi\'c-Jelisav\v{c}i\'c, M. Pandurovi\'c and I. Smiljani\'c\\
\llap{}Vin\v{c}a Institute of Nuclear Sciences, University of Belgrade,\\
  P.O. Box 522, 11001 Belgrade, Serbia\\
  E-mail: \email{slukic@vinca.rs}}

\preprint{LCD-Note-2012-008, arXiv:1301.1449}

\abstract{
Procedures for correcting the beam-beam effects in luminosity measurements at CLIC at 3 TeV center-of-mass energy are described and tested using Monte Carlo simulations. The angular counting loss due to the combined Beamstrahlung and initial-state radiation effects is corrected based on the reconstructed velocity of the collision frame of the Bhabha scattering. The distortion of the luminosity spectrum due to the initial-state radiation is corrected by deconvolution. At the end, the counting bias due to the finite calorimeter energy resolution is numerically corrected. To test the procedures, BHLUMI Bhabha event generator, and Guinea-Pig beam-beam simulation were used to generate the outgoing momenta of Bhabha particles in the bunch collisions at CLIC. The systematic effects of the beam-beam interaction on the luminosity measurement are corrected with precision of 1.4 permille in the upper 5\% of the energy, and 2.7 permille in the range between 80 and 90\% of the nominal center-of-mass energy.
}


\begin{document}
%
\section{Introduction}

The future linear colliders CLIC \cite{CLIC} and ILC \cite{ILC} are designed for precision measurements in elementary-particle physics, complementing measurements performed at the Large Hadron Collider (LHC) now operating at CERN. Despite significant differences in accelerator technology, as well as in center of mass (CM) energy and charge density, the detector design is to a large extent common to both projects \cite{CDR12, ILD}. This is, in particular, true for the instrumentation of the forward region of the detector, including the luminosity calorimeter LumiCal \cite{Abr10, Abr09}. The present study is part of this common effort, and is applicable with small differences in both contexts. The results for ILC have been reported elsewhere \cite{Luk12b}.

Luminosity, $L$, and luminosity spectrum, $\mathcal{L}(E_{CM})$\footnote{The precise definition of the term "luminosity spectrum" as used in this work is given in section \ref{sec-phys}}, are key inputs to many measurements at collider experiments, including mass and cross-section measurements, as well as production-threshold scans. Precision of the luminosity measurement is critical at linear colliders in order to match the inherent precision potential of the lepton machines. 
The most precise luminosity measurement method at linear colliders to date is
to count Bhabha-scattering events recognized by coincident detection of showers in the fiducial volume (FV) in both halves of the luminometer in the very forward region in a given energy range. The number of events, $N$, is then divided by the Bhabha cross section, $\sigma$, integrated in the corresponding region of the phase space. Bhabha scattering is a well-known QED process and at several experiments at LEP this technique allowed reaching sub-permille precision \cite{Opal00, Aleph00, L3_00, Arb96, Jad03}. At future linear colliders, however, CM energy will be 3 to 30 times higher, and instantaneous luminosity up to three orders of magnitude higher \cite{CLIC, ILC}. At such high beam power density, the energies and the polar angles of the Bhabha particles are strongly influenced by beam-beam effects \cite{Yok91, Sch96}, which creates severe Bhabha counting losses. 

The expression for measured luminosity can be formally written as follows,

\begin{equation}
\label{eq-luminosity}
L = \frac{N(\Xi(\Omega^{lab}_{1,2}, E^{lab}_{1,2}))}{\sigma(Z(\Omega^{CM}_{1,2}, E^{CM}_{1,2}))},
\end{equation}

Here $\Xi(\Omega^{lab}_{1,2}, E^{lab}_{1,2})$ is a function describing the selection criteria for counting the detected events based on the angles $\Omega^{lab}_{1,2}$ and energies $E^{lab}_{1,2}$ of the final particles in the lab frame, and $Z(\Omega^{CM}_{1,2}, E^{CM}_{1,2})$ is a function describing the corresponding region of phase space where the cross section is integrated. These functions can be expressed as products $\Xi = \prod_i \xi_i$ and $Z = \prod_i \zeta_i$ where the functions $\xi_i$ and $\zeta_i$ are based on specific topological and kinematical properties of the detected/generated pair. For each $i$, the physical meaning of $\xi_i$ and $\zeta_i$ corresponds to each other, although their mathematical form may be different (See in particular section \ref{sec-BS} and Eqs. \ref{eq-xi-ang} and \ref{eq-zeta-ang}). The set of functions $\xi_i$ and $\zeta_i$ includes the angular selection requiring both particles to be detected in the FV, as well as the energy range selection and possible further cuts to eliminate background.

Because of the random and asymmetric momentum loss when electrons emit Beamstrahlung, the CM frame of the Bhabha process moves with respect to the lab frame with axial velocity different for every colliding pair. As a consequence, $\Xi$ and $Z$ operate on kinematical arguments in different reference frames. Thus, if $\Xi$ and $Z$ have the same form, different regions of the phase space will be covered, leading to a systematic bias in the luminosity measurement. This systematic bias cannot be neglected at the future linear colliders, and is particularly accute at the 3 TeV CLIC \cite{CLIC}. 

A way around this problem is to define $\Xi$ and $Z$ such that the counting rate is independent of the reference frame. Some of the functions $\xi_i$ and $\zeta_i$ can be defined invariant to the boost along the beam axis. This is, for example, the case with the cuts on the reconstructed CM energy. However, the requirement that the outgoing particles hit the FV of the detector on both sides does not possess such invariance. In this paper, a definition of $\xi_{FV}$ and $\zeta_{FV}$ is proposed such that both the experimental count $N$ and the cross-section $\sigma$ are reconstructed in the same reference frame, namely the collision frame, which will be defined in section \ref{sec-phys}. 

The physical processes affecting the luminosity measurement are outlined and the used terms and notation defined in section \ref{sec-phys}. The analysis method with the correction procedures, as well as the test results are described in section \ref{sec-corr}. In the conclusions, the main advantages of the presented method are restated, and the final uncertainties are listed and briefly discussed.

\section{The physical processes affecting the luminosity measurement and an outline of the correction procedure}
\label{sec-phys}

\begin{figure}[ht]
\centering
\includegraphics{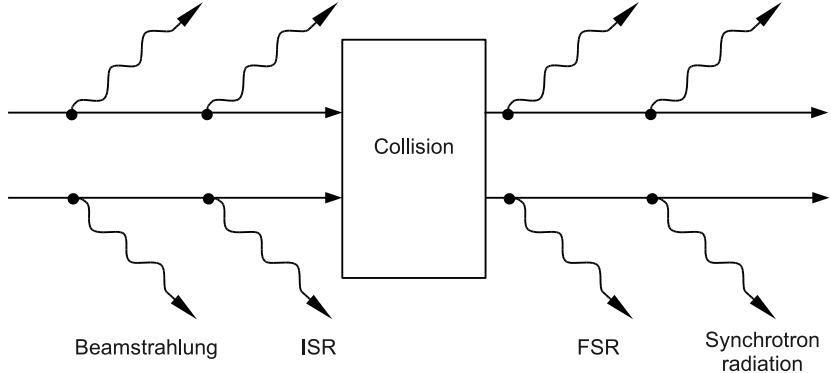}
\caption{\label{fig-scattering}Schematic representation of the physical processes affecting the luminosity measurement}
\end{figure}

The sequence of physical processes relevant to the present discussion is schematically represented in figure \ref{fig-scattering}. Due to the pinch effect during the bunch collision, both particles may emit Beamstrahlung photons and so lose energy and momentum before the interaction. Thus in general, $E_{CM} < E_0 \equiv 2 E_{beam}$. The CM energy distribution at this stage is the actual luminosity spectrum $\mathcal{L}(E_{CM})$. The probability of the Bhabha scattering scales with $1/s \equiv 1/E^2_{CM}$, resulting in the CM energy distribution of the Bhabha events $\mathcal{B}(E_{CM}) \propto \mathcal{L}(E_{CM})/E^2_{CM}$. The Bhabha process is itself accompanied by emission of initial-state radiation (ISR) that is nearly collinear with the initial particle momenta, as well as final-state radiation (FSR) that is approximately collinear with the outgoing particle momenta. Since ISR is nearly collinear with the beam axis, it misses the luminometer, so that the CM energy reconstructed from the detected particles is $E_{CM,rec} < E_{CM}$, and the corresponding spectrum is,

\begin{equation}
\label{eq-ecm-det}
h(E_{CM,rec}) = \int\limits_0^{E_{\max}}{\mathcal{B}(E_{CM}) \frac{1}{E_{CM}} \mathcal{I}(\frac{E_{CM,rec}}{E_{CM}}) \ud E_{CM}}
\end{equation}

where $\mathcal{I}(x)$ is the distribution of the fractional CM energy losses due to ISR. $\mathcal{I}(x)$ is approximately independent of $E_{CM}$.

Due to the finite energy resolution of the luminometer, the reconstructed spectrum is smeared, which can be represented as a convolution with a normalized Gaussian\footnote{Strictly speaking, the smearing width depends on the deposited energy of the showers. However, as only a relatively narrow energy range is being analyzed here, the smearing width will be treated as being approximately constant.}.

\begin{equation}
\label{eq-ecm-rec}
h^*(E_{CM,rec}) = \frac{1}{\sqrt{2\pi} \sigma} \int\limits_{0}^{\infty} h(E'_{CM,rec}) \exp \left(-\frac{(E_{CM,rec}-E'_{CM,rec})^2}{2 \sigma^2} \right) \ud E'_{CM,rec}
\end{equation}

The term \emph{collision frame} will be used here for the frame of the two-electron system\footnote{Unless stated otherwise, electron always refers to electron or positron} after emission of Beamstrahlung and ISR and before emission of FSR\footnote{In reality, ISR and FSR can not be cleanly separated even theoretically, due to the quantum interference between them. Thus in practice the collision frame is defined as the CM frame of the final electrons together with all radiation within a given tolerance angle with respect to the final electron momenta. The assumption of clean separation between ISR and FSR introduces a small uncertainty in the final result.}. The scattering angle in the collision frame is denoted $\theta^{coll}$. Due to the radiation prior to the collision, the collision frame has a non-zero velocity $\vec\beta _{coll}$, and the outgoing particle angles in the lab frame, $\theta^{lab}_1$ and $\theta^{lab}_2$, are not symmetric. In a significant fraction of events, the acollinearity is so large that the two particles are not detected in coincidence within the FV of the luminometer. In this way, Beamstrahlung and ISR induce an \emph{angular counting loss} of Bhabha events.

Finally the electromagnetic deflection (EMD) of the outgoing electrons in the field of the opposing bunch induces a small additional angular counting loss.

The outline of the procedure of the Bhabha-count analysis is as follows:
\begin{enumerate}
\item Reconstruct the CM energy $E_{CM,rec}$ and the collision-frame velocity $\beta_{coll}$ for each pair detected in the FV of the luminometer, from the angles and the measured particle energies.
\item Assign weights to events to correct for the acceptance reduction due to $\vec\beta _{coll}$, as shown in section \ref{sec-BS}. 
\item Deconvolution of the ISR energy loss $\mathcal{I}(x)$ from the spectrum $h^*(E_{CM,rec})$, in order to restore the $\mathcal{B}^*(E_{CM})$ CM energy spectrum of the Bhabha events (section \ref{sec-deconvolution}).
\item Integrate $\mathcal{B}^*(E_{CM})$ over the energy range of measurement. 
\item Correct the systematic effect of the finite energy resolution of the luminometer on the number of counts in the peak (section \ref{sec-Eresp}).
\end{enumerate}

The absolute luminosity in the measured energy range is then given by equation \ref{eq-luminosity}, and the approximate differential form of the luminosity spectrum with the luminometer energy smearing can be obtained as $\mathcal{L}^*(E_{CM}) \propto \mathcal{B}^*(E_{CM}) E^2_{CM}$. In the following section the precision of different correction steps will be tested by MC simulation, and expressed as relative contribution to the luminosity uncertainty $\Delta L_\alpha/L$ for each step $\alpha$.

\section{Analysis and correction procedures}
\label{sec-corr}

\subsection{Simulation methods used to test the analysis procedure}
\label{sec-sim}

To test the analysis procedure, Bhabha events in the bunch-collision were simulated with the Guinea-Pig software \cite{Sch96}. The initial bunch coordinate- and momentum distributions were taken from the simulation results by D. Schulte et al. \cite{Sch02}. The coordinate distribution covered more than 10 $\sigma$ bunch widths both in the horizontal and the vertical directions. The angular distribution of the particles in the bunch was quasi-Gaussian, with transverse emittance of 660 nm rad in the horizontal, and 20 nm rad in the vertical direction. The bunch collision was simulated in the CM frame of the colliding bunches, which is equivalent to a head-on collision with zero crossing angle. The beam overlap reduction due to the crossing angle is offset by the crab-crossing scheme. 

The Bhabha events were produced using a method resembling that used by C. Rimbault et al. \cite{Rim07}: 
\begin{itemize}
\item The initial four-momenta of the colliding $e^- e^+$ pairs are generated in Guinea-Pig by beam-beam simulation 
\item The decision is made whether the Bhabha scattering will be realized in the collision, based on the $1/s$ proportionality of the Bhabha cross section.
\item If a Bhabha event is to be realized, the final four-momenta are picked from a file generated at 3 TeV by the BHLUMI generator \cite{Jad97}.  
\item The final momenta are scaled to the CM energy of the colliding pair, rotated to match the collision axis, and boosted back to the lab frame. 
\item Finally the outgoing Bhabha electrons are tracked to simulate the electromagnetic deflection. 
\end{itemize}

Nearly four million Bhabha events were generated. As the polar angles of the two electrons are often severely shifted in the opposite directions when the momenta are boosted into the lab frame, the polar-angle cuts in the generator frame were kept very wide - between 10 and 200 mrad. On the other hand, in order to avoid simulating a large number of events with very low scattering angles, post-generator cuts were applied in the collision frame, and only events with the scattering angle between 37 and 90 mrad were kept in the file. As the limiting angles of the luminometer FV at CLIC are 43 and 80 mrad \cite{Schw11}, these cuts leave a safety margin of 6, respectively 10 mrad, to accomodate for the small parallel shift of the polar angles that can be induced by the EMD and by the off-axis ISR.

The interaction with the detector was approximated in the following way:
\begin{itemize}
\item The four-momenta of all electrons and photons within 5 mrad of the most energetic shower were summed together on each side. The 5 mrad criterion corresponds closely to the Moli\`ere radius of the high-energy showers in the luminometer \cite{Sad08}. The resulting four-momenta were taken to represent the detected final particles. 
\item The energy resolution of the luminometer was included by adding random Gaussian fluctuations to the final particle energies. The standard deviation of energy was parametrized as $\sigma_E / E = \sqrt{a^2/E + b^2}$. The value of the stochastic term is $a = 0.21$ in all relevant analyses \cite{Schw11, CDR12, Agu11}. The constant term $b$ is zero in Ref. \cite{Schw11}, and 1.1\% in Ref. \cite{Agu11}. The correction procedure was tested with three different values of the constant term $b$: 0, 0.35\% and 1.1\%. The results of these three tests agree within their respective statistical uncertainties. Only results for $b=0$ are presented in this paper.
\item The finite angular resolution of the luminometer was included by adding random fluctuations to the final particle polar angles. The nominal value of $\sigma_\theta = 2.2 \times 10^{-5} \text{\, rad}$ estimated for the ILC version of LumiCal \cite{Sad08} was used. Higher values for $\sigma_\theta$ were also tested, but no significant effect on the final uncertainties was found for $\sigma_\theta < 2 \times 10^{-4} \text{\, rad}$.
\end{itemize}

\subsection{Invariant counting in the collision frame}
\label{sec-BS}

The movement of the collision frame with respect to the lab frame is responsible for the acollinearity leading to the angular counting loss. The velocity of the collision frame with respect to the lab frame $\vec{\beta} _{coll}$, can be calculated from the measured polar angles. If $\beta_{coll}$ is taken to be collinear with the $z$-axis, the expressions for the boost of the Bhabha scattering angles into the lab frame give,

\begin{equation}
\label{eq-beta}
\beta_{coll} = \frac{\sin (\theta^{lab}_1 + \theta^{lab}_2)}{ \sin \theta^{lab}_1 + \sin \theta^{lab}_2 }
\end{equation}

Equation \ref{eq-beta} does not depend on any assumptions about the number of emitted ISR and Beamstrahlung photons, nor on their direction, apart from the assumption that the vector sum of their momenta is collinear with the z-axis\footnote{\label{foot-radial} Strictly speaking, $\vec{\beta} _{coll}$ has a small radial component $\beta_\rho$, which is larger than 0.01 in only about 5 permille of cases. However, the influence of $\beta_\rho$ on the polar angles of the Bhabha pair is almost indistinguishable from an additional axial boost. Thus for the purpose of recovering the counting loss due to the acollinearity, $\beta_{coll}$ is approximately treated as a scalar quantity.}.

\begin{figure}[ht]
\centering
\includegraphics{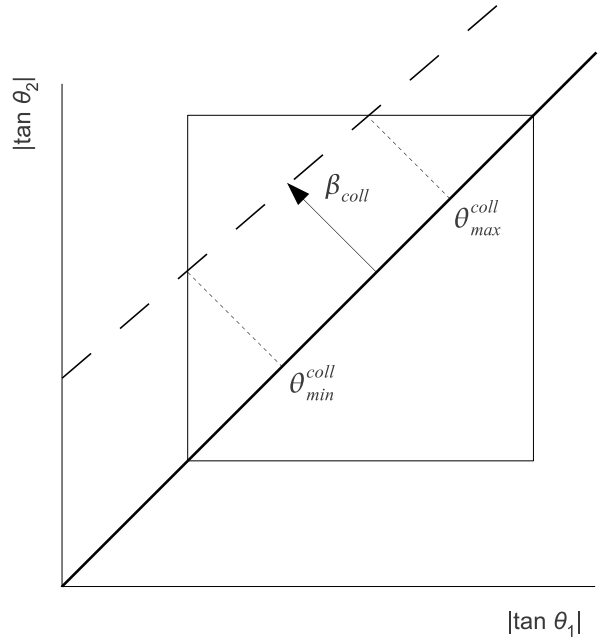}
\caption{\label{fig-asymmetry}Schematic representation of the distortion of the polar angles due to the movement of the collision frame. The box represents the region in which both electrons hit the FV, and the dashed line represents the event subset characterized by a given $\beta_{coll}$. $\theta^{coll}_{\min}$ and $\theta^{coll}_{\max}$ denote the effective limiting scattering angles for this subset.}
\end{figure}

If events from a subset characterized by a given $\beta_{coll}$ are plotted in the $|\tan \theta_2|$ vs. $|\tan \theta_1|$ graph, they lie on a line displaced from the central diagonal, as schematically represented by the dashed line in figure \ref{fig-asymmetry}. As can be seen from the figure, the range of accepted scattering angles decreases with increasing $\beta_{coll}$. The effective limiting angles $\theta^{coll}_{\min}$ and $\theta^{coll}_{\max}$ for the subset of events charaterized by a given $\beta_{coll}$ are obtained by boosting $\theta_{\min}$ and $\theta_{\max}$ into the collision frame.

To account for the smaller acceptance of the events characterized by a given $\beta_{coll}$, every event has to be weighted with the appropriate correction factor. In this way, the number of events between $\theta_{\min}$ and $\theta_{\max}$ in the collision frame is recovered for each $\beta_{coll}$ subset separately. The weighting factor is defined as,

\begin{equation}
\label{eq-w}
w(\beta_{coll}) = \frac{\int\limits^{\theta_{\max}}_{\theta_{\min}} \frac{\ud\sigma}{\ud\theta} \ud\theta }{\int\limits^{\theta^{coll}_{\max}}_{\theta^{coll}_{\min}} \frac{\ud\sigma}{\ud\theta} \ud\theta}. 
\end{equation}

The FV selection function is thus defined as\footnote{By the standard definition of the polar angle $\theta$, the interval corresponding to the FV on the forward side of the IP is $(\theta_{\min}, \theta_{\max})$, and on the backward side, $(\pi - \theta_{\max}, \pi - \theta_{\min})$},

\begin{equation}
\label{eq-xi-ang}
\xi_{FV} = \left\{ 
   \begin{array}{cl}
      w & ; \theta^{lab}_{1,2} \in FV \\
      0 & ; otherwise \\
   \end{array} \right.
\end{equation}

Using this FV selection function, the number of events $N$ satisfying the condition $\theta^{coll} \in (\theta_{\min}, \theta_{\max})$ in the collision frame is reconstructed. The corresponding function $\zeta_{FV}$ for the cross-section integration is thus,

\begin{equation}
\label{eq-zeta-ang}
\zeta_{FV} = \left\{ 
   \begin{array}{cl}
      1 & ; \theta^{coll} \in (\theta_{\min}, \theta_{\max}) \\
      0 & ; otherwise \\
   \end{array} \right.
\end{equation}

\subsubsection{Test of the collision-frame counting method}

To test the counting method, histograms of $E_{CM,rec}$ reconstructed from kinematic parameters of the detected particles were generated as follows: 

\begin{description}
\item [Control histogram]: All events with the scattering angle in the collision frame $\theta^{coll}$ such that $\theta_{\min} < \theta^{coll} < \theta_{\max}$ are accepted. Therefore this histogram is not affected by counting losses due to Beamstrahlung and ISR. This is, of course, only possible in the simulation.
\item [Uncorrected histogram]: Events hitting the FV of the luminometer in the lab frame.
\item [Corrected histogram]: Events hitting the FV of the luminometer in the lab frame, stored with the weight $w$ calculated according to equation \ref{eq-w} 
\end{description}

\begin{figure}[b]
\centering
\includegraphics{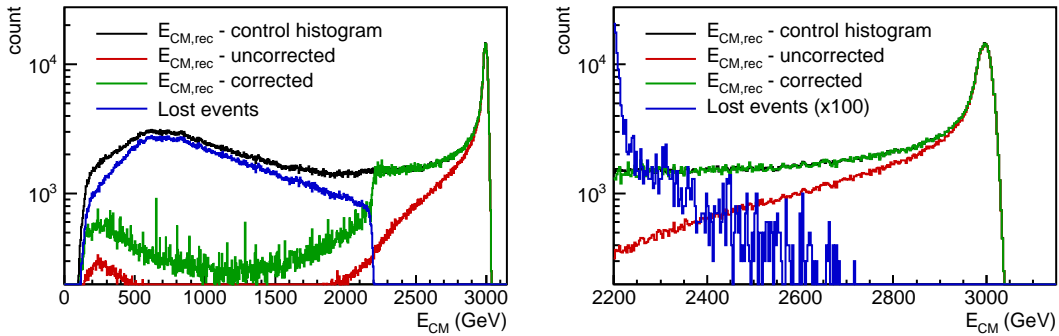}
\caption{\label{fig-BS-corr}Correction of the counting loss due to Beamstrahlung and ISR. Left: whole spectrum; right: zoom on energies above 2200 GeV. Black: Simulated control spectrum without counting loss due to Beamstrahlung and ISR; red: Reconstructed $E_{CM}$ spectrum affected by the counting loss; green: Reconstructed spectrum with correction for the counting loss due to Beamstrahlung and ISR; blue: events inaccessible to the correction due to high $\beta_{coll}$ (see text).}
\end{figure}

The full kinematical information, including the energy of the detected final particle, was used for the reconstruction of the CM energy. To calculate the correction weight $w$, the approximate expression for the angular differential cross section $d\sigma / d\theta \approx \theta^{-3}$ was used. The results are shown in figure \ref{fig-BS-corr}. The control spectrum is plotted in black, red is the spectrum affected by the counting loss, green is the corrected spectrum. 

The blue line in figure \ref{fig-BS-corr} represents the events inaccessible to the correction due to their high values of $\beta_{coll}$. In the subsets of events characterized by $\beta_{coll}$ above a certain treshold, at least one electron is always lost (see figure \ref{fig-asymmetry}). However, for such events, the Beamstrahlung-ISR energy loss is also above a certain minimum, so that they are only present in significant number below 2200 GeV. A small number of high-$\beta_{coll}$ events are also present at energies above 2200 GeV, as seen in the zoomed figure on the right, where these events are scaled by a factor 100. In these events, $\vec{\beta} _{coll}$ has a relatively high radial component, due to the off-axis radiation before collision. This increases the acollinearity of such events relative to other events with similar energy loss (see footnote \ref{foot-radial}). The relative contribution of these events to the peak integral above 95\% of the nominal CM energy is of the order of $2\times 10^{-5}$.

Before correction, the counting loss in the peak integral above 95\% of the nominal CM energy was 3.8\%. After correction, the remaining relative deviation in the peak with respect to the control spectrum is $(-0.1 \pm 0.4 (\text{stat.})) \times 10^{-3}$. In the tail between 80\% and 90\% of the nominal CM energy, the counting loss before correction was 43.1\%. After correction, the remaining relative deviation in the tail is $(-3.6 \pm 1.8 (\text{stat.})) \times 10^{-3}$, which includes a deviation of $(-2.7 \pm 0.1) \times 10^{-3}$ due to the lost events. The statistical uncertainty of the remaining deviation was estimated taking into account the correlations between the corrected and the control spectra. The precision of the Beamstrahlung-ISR correction is of the order of permille despite the presence of the following sources of systematic uncertainty of the correction:

\begin{itemize}
\item The assumption that the deformation of the Bhabha angles induced by Beamstrahlung and ISR is well described as a Lorentz boost along the beam axis (this assumption is the source of the "lost" events in the peak), 
\item The implicit assumption that the cluster around the most energetic shower always contains the Bhabha electron. In a small fraction of events, this is not the case, and the reconstructed polar angles $\theta_{1,2}^{lab}$ do not correspond to the final electron angles.
\item The use of the approximate angular differential cross section for the Bhabha scattering in the calculation of $w$,
\item Assumption that all ISR is lost, and all FSR is detected (this assumption has, in principle, an influence on the calculation of $\beta_{coll}$, and consequently on $w$). 
\end{itemize}

\subsection{Deconvolution of the ISR energy loss}
\label{sec-deconvolution}

After correcting for the angular counting loss, the ISR energy loss can be deconvoluted from the resulting spectrum $h(E_{CM,rec})$ to restore the CM energy spectrum of the Bhabha events $\mathcal{B}^*(E_{CM})$\footnote{The Bhabha-event spectrum is marked with a star here, because it is smeared by the finite energy resolution of the luminometer. See section \ref{sec-Eresp}.}. When data according to equation \ref{eq-ecm-det} is binned in $N$ sufficiently narrow bins, it takes approximately the discrete form,

\begin{equation}
\label{eq-h-disc}
h_i \approx \sum\limits_{j=1}^N \mathcal{I}_{ij} \mathcal{B}^*_j
\end{equation}

As the $\mathcal{I}_{ij}$ matrix has triangular form, equation \ref{eq-h-disc} can be solved for $\mathcal{B}^*_j$ exactly, using the Jacobi method. The solution proceeds from high-energy towards the lower-energy bins, indroducing an increasing uncertainty towards lower energies. 

To obtain $\mathcal{I}_{i,j}$, the function $\mathcal{I}(x)$ was parametrized by the beta distribution used for the parametrization of the beam spectra of linear colliders \cite{Ohl97},

\begin{equation}
\label{eq-g}
\mathcal{I}(x) = a_0 \delta(x-1) + \left\{ 
   \begin{array}{cl}
      a_1 x^{a_2} (1-x)^{a_3} & ; x<1 \\
      0                       & ; x \ge 1 \\
   \end{array} \right.
\end{equation}

\begin{figure}[ht]
\centering
\includegraphics{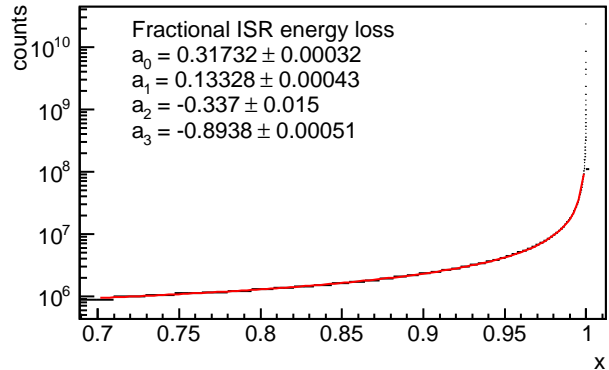}
\caption{\label{fig-isr}Fit of the relative energy-loss distribution due to ISR.}
\end{figure}

The parameters were obtained by fitting equation \ref{eq-g} to the fractional CM energy distribution after ISR, reconstructed from the same BHLUMI data set as used in Guinea-Pig. The fit was performed with variable binning in order to have sufficiently fine binning near $x=1$, while avoiding large differences in statistical uncertainties for individual bins. The data histogram was first normalized to the unit integral. The results are shown in figure \ref{fig-isr}. The parameter $a_0$ was obtained as the ratio of the number of counts in the narrow peak above $x = 0.99995$ to the number of counts in the entire spectrum, and the remaining coefficients were obtained by fitting the function to the data in the range (0.7, 0.99995)\footnote{\label{ft-cut} The angular generator cuts in the lab frame cause significant losses in the distribution for $x<0.5$ because high energy loss in ISR emission correlates with high acollinearity. This affects the overall normalization, and thus the value of $a_0$. The value of $a_0$ obtained here is appropriate for the deconvolution of the simulated spectrum where the same set of BHWIDE samples was used. However, for the analysis of the real experimental data, ideally the distribution without cuts in the lab frame should be used.}\footnote{The functional form of equation \ref{eq-g} suggests that the ratio $a_0/a_1$ can be fixed by the normalization requirement. However, the beta distribution fails to properly describe the form of $\mathcal{I}(x)$ for $x < 0.7$ (regardless of the angular cuts in the lab frame discussed above), so that the overall norm is different than the integral of the beta distribution extrapolated from the fit. Therefore, $a_1$ was allowed to vary freely in the fit.}.


\subsubsection{Test of the deconvolution procedure}

\begin{figure}
\centering
\includegraphics{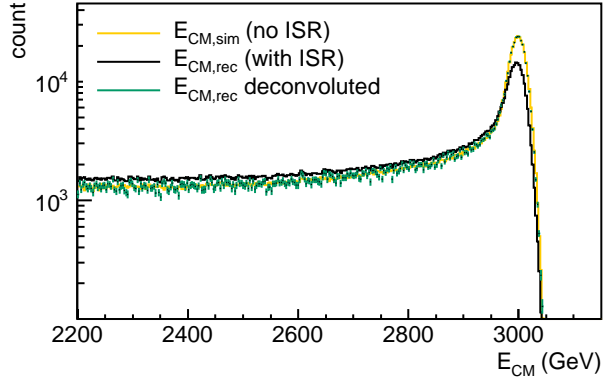}
\caption{\label{fig-deconv}Deconvolution of the ISR deformation of the luminosity spectrum. Yellow: the control histogram -- simulated $E_{CM}$ before emission of ISR, smeared with a normalized Gaussian; black: the histogram affected by the ISR energy loss -- reconstructed $E_{CM}$ from the detected showers, green: deconvoluted spectrum}
\end{figure}

In this test, the following histograms were generated:

\begin{description}
\item[Control histogram] was filled with simulated CM energies before ISR emission, and then smeared with a normalized Gaussian with constant width corresponding to the luminometer energy-resolution at the peak energy. 
\item[Histogram with ISR energy loss] $h(E_{CM,rec})$ is the same as the control histogram from section \ref{sec-BS} -- filled with energies reconstructed from the final-state kinematics, and with inclusion of the luminometer energy resolution. 
\item[Deconvoluted histogram] was obtained by solving the system of linear equations represented by equation \ref{eq-h-disc}, taking the binned data of the affected histogram as $h_j$.
\end{description}
For each histogram, event selection was made on the scattering angles in the collision frame, so that the Beamstrahlung-ISR angular counting loss is not present. This was done in order to assess the accuracy of the deconvolution separately from the Beamstrahlung-ISR counting-loss correction. Results are shown in figure \ref{fig-deconv}. 

Before deconvolution, the relative counting loss in the peak above 95\% of the nominal CM energy was 23.4\%. After deconvolution, the relative remaining deviation of the peak integral with respect to the control histograms is $(+1.3 \pm 2.1) \times 10^{-3}$. In the tail between 80\% and 90\% of the nominal CM energy, the ISR energy loss increases the count by 14.5\%. After deconvolution, the remaining deviation in the tail is $(-2.3 \pm 3.9) \times 10^{-3}$. 

The contributions from the uncertainties of the fitted parameters of the ISR energy-loss function $\mathcal{I}(x)$ were added to the statistical uncertainty of the remaining deviation after deconvolution. The full covariance matrix of the fit parameters was used, together with the partial derivatives of the count estimated by variation of the fit parameters by one sigma, one parameter at a time. With the statistics of about four million generated Bhabha events, the uncertainties due to the fit parameters are $(\Delta N/N)_{peak,ISRfit} = 0.53 \times 10^{-3}$ for the peak, and $(\Delta N/N)_{tail,ISRfit} = 0.07 \times 10^{-3}$. 


\subsection{Effect of the luminometer energy resolution on the counting rate in the peak}
\label{sec-Eresp}

The finite energy resolution of the luminometer introduces a counting bias in two ways:
\begin{enumerate}
\item \label{slope-cut} By asymmetric redistribution of events from each side of the sharp energy cut $E_{cut}$ used to define the energy range, due to the slope of the underlying distribution at the position of the cut.
\item \label{tail-cut} By smearing the luminosity peak so that a small part of it is cut off below $E_{cut}$.
\end{enumerate}

The second effect is difficult to precisely correct because of the strong dependence on the position of the energy cut, and because of the uncertainties of the inherent width of the luminosity peak and of the energy resolution, as well as the strong correlations between the fitted parameters that dominate the spectrum in the peak area (see Eqs. \ref{eq-lumispec} and \ref{eq-F-param}). However, if the energy cut is made at a sufficient distance from the peak, the second effect becomes negligible, and the energy-resolution effect can be precisely corrected based on the parametrization of the functional form of the experimental spectrum after deconvolution of ISR.

\begin{equation}
\label{eq-lumispec}
\mathcal{B}^*(E_{CM}) = \frac{1}{\sigma \sqrt{2 \pi}} \int\limits_{0}^{\infty} \mathcal{B}(E') \exp \left(- \frac{(E_{CM}-E')^2 }{ 2 \sigma^2 } \right) \ud E'
\end{equation}

If the inherent width of the luminosity peak is neglected, $\mathcal{B}(E_{CM})$ can be parametrized by the beta distribution,

\begin{equation}
\label{eq-F-param}
\mathcal{B}(E_{CM}) = b_0 \delta(E_{CM}-E_0) + \left\{ 
   \begin{array}{cl}
      b_1 E_{CM}^{b_2} (E_0-E_{CM})^{b_3} & ; E_{CM}<E_0 \\
                  0                       & ; E_{CM} \ge E_0 \\
   \end{array} \right.
\end{equation}

One may recall here that the use of a constant standard deviation $\sigma$ in equation \ref{eq-lumispec} is an approximation, as $\sigma$ depends on the particle energy, and is thus different for different $E_{CM}$. The systematic error induced by the energy resolution of the luminometer can now be expressed as,

\begin{equation}
\label{eq-err-Eres}
\frac{\Delta N_{Eres}}{N} = \frac{ \int\limits_{E_{cut}}^{ \infty } \frac{E_{CM}^2}{E_0^2} (\mathcal{B}^*(E_{CM}) - \mathcal{B}(E_{CM})) \ud E_{CM} }{\int\limits_{E_{cut}}^{ \infty } \frac{E_{CM}^2}{E_0^2} \mathcal{B}(E_{CM}) \ud E_{CM} }
\end{equation}

This expression can now be estimated by numerical integration based on the fitted parameters of $\mathcal{B}^*(E_{CM})$ (Eqs. \ref{eq-lumispec} and \ref{eq-F-param}). Even though the reproduction of the integral count by integration of the fitted function has in principle limited accuracy, rather accurate prediction of the relative error (equation \ref{eq-err-Eres}) is achieved. The fit was performed on the deconvoluted histogram with the fixed parameters $E_0 = 3 \text{ TeV}$ and $\sigma = 13.7 \text{ GeV}$, while $b_{0-3}$ were varying freely. The value of $\sigma$ was obtained by fitting the data in the region of the luminosity peak. It contains contributions from both the energy resolution of the luminometer and the inherent width of the peak. Neglecting the inherent width introduces an uncertainty of 20\% in the magnitude of the correction. This represents a conservative estimate of the precision with which $\sigma$ can be known and, as shown below, the results obtained with this assumption are acceptable.

\begin{figure}
\centering
\includegraphics{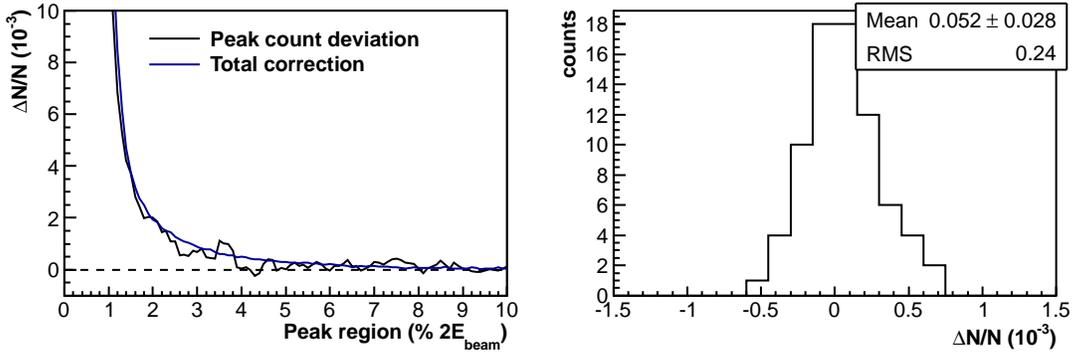}
\caption{\label{fig-Eresp}Left: relative deviation of the peak count induced by the luminometer energy resolution in the reconstructed spectrum as a function of the peak region expressed as fraction of the nominal CM energy $E_0$, compared to the predicted value based on the fitted spectrum. Right: histogram of the normalized remaining deviations of the peak count after correction (see text), for $E_{cut}$ more than 2.5\% away from $E_0$.}
\end{figure}

The relative deviation of the count in the reconstructed peak is shown in the left pane of figure \ref{fig-Eresp} as a function of the relative distance of the energy cut to the peak energy in percent (black line). The predicted deviation according to equation \ref{eq-err-Eres} is also shown for comparison (blue line). There is an excellent agreement between the predicted and the simulated deviations.
To take a safe distance from the peak, only points for which $E_{cut}$ is more than 2.5\% away from $E_0$, corresponding to about 5 $\sigma$ of the fitted peak, will be considered in the following. 

The fluctuations of the simulated deviation curve in figure \ref{fig-Eresp} are of statistical nature. These fluctuations can be used as an external measure of the statistical uncertainty of the counting bias in the simulation. In the right pane in figure \ref{fig-Eresp}, the histogram of these fluctuations is shown, calculated as remaining deviations after correction, for $E_{cut}$ more than 2.5\% away from $E_0$. The RMS of the fluctuations corresponds to a relative statistical uncertainty of $0.24 \times 10^{-3}$ with respect to the the peak count in the top 5\%. The relative deviation in the top 5\% estimated from equation \ref{eq-err-Eres} is $-0.29 \times 10^{-3}$. The mean remaining bias after correction is $(0.05 \pm 0.03) \times 10^{-3}$.

Similar procedure was applied to estimate the relative bias and the remaining uncertainty in the tail region from 80\% to 90\% of $E_0$. The RMS of the fluctuations is $0.79 \times 10^{-3}$, the uncorrected deviation is $+0.32 \times 10^{-3}$, and the remaining deviation after correction is $(0.09 \pm 0.09) \times 10^{-3}$.

\subsection{The Electromagnetic Deflection}
\label{sec-EMD}

To estimate the counting loss due to the EMD, the angular selection was applied once before and once after the deflection in the simulation, and the relative difference in the resulting number of events was calculated. The EMD counting loss above 95\% of the nominal CM energy is $(-0.50 \pm 0.05) \times 10^{-3}$. In the tail from 80 to 90\% of the nominal CM energy, the EMD counting loss is $(-1.08 \pm 0.08) \times 10^{-3}$. 

\section{Conclusions}

A method of invariant counting of Bhabha events was presented. The number of Bhabha events within a given range of scattering angles in the collision frame, and in a given range of $E_{CM}$ is reconstructed. The corresponding limits can be used for the cross-section integration in a straightforward way. In this way the luminosity expression (equation \ref{eq-luminosity}) is essentially insensitive to the beam-beam effects.


The remaining systematic uncertainties of the Bhabha count with the presented methods were estimated by MC simulations. In addition, the systematic uncertainty due to the EMD-induced counting loss was estimated and found to be small. The remaining relative errors in the top 5\%, as well as in the tail from 80 to 90\% of the nominal CM energy are listed in table \ref{tab-unc}. Beam-beam effects in the luminosity measurement at 3 TeV CLIC can be corrected and the luminosity spectrum reconstructed with a few permille precision above ca. 73\% of the nominal CM energy. 


\begin{table}[ht]
\caption{\label{tab-unc}Relative remaining error after correction of different systematic effects in luminosity measurement in the peak above 95\% and the tail from 80 to 90\% of the nominal CM energy. The fraction of events with high $\beta_{coll}$ constitutes part of the remaining bias of the Beamstrahlung-ISR angular loss correction. The last column gives the total remaining error when the high $\beta_{coll}$ contribution is corrected.}
\begin{center}
\begin{threeparttable}
\begin{tabular}{@{} c l  R-{2}{3} @{$\pm$} l   R-{2}{2} @{$\pm$} l  @{}}
\# & Effect & \multicolumn{2}{c}{Top 5\%} & \multicolumn{2}{c}{80 - 90\% of $E_0$} \\
   & & \multicolumn{2}{c}{($10^{-3}$)} & \multicolumn{2}{c}{($10^{-3}$)} \\
\hline
1 & Beamstrahlung-ISR angular loss          & -0.1 &0.4   &  -3.6&1.8  \\
2 & High $\beta_{coll}$ \tnote{a}           & -0.019&0.008 & -2.7&0.1  \\
3 & ISR energy-loss                         & +1.3 &2.0   &  -2.3&3.9   \\
4 & Energy resolution                       &  0.05&0.03  &   0.09&0.09   \\
5 & EMD counting loss (uncorrected)         & -0.50&0.05  &  -1.08&0.08    \\
\hline
  & Total                                   &  1.4  &2.0  &   4.4 &4.3  \\
  & Total (corrected for \#2)               &  1.4  &2.0  &   2.7 &4.3  \\
\hline
\end{tabular}
\end{threeparttable}
\end{center}
\end{table}

In a separate study for the case of ILC \cite{Luk12b}, it was shown that the method presented here is robust with respect to unaccounted-for bunch size and charge variations up to 20\%, as well as the vertical and horizontal offset up to 1 $\sigma$ bunch height and width, respectively.

%
%
\section{Acknowledgements}

This work has been funded by the Ministry of science and education of the Republic of Serbia under the project Nr. OI 171012, "Physics and detector studies in HEP experiments". Additional travel grants from the CLIC Physics and Detectors study are gratefully acknowledged. Thanks to A. Sailer and D. Schlatter for many useful remarks during the review of the LCD note.

%
\bibliographystyle{JHEP}
\bibliography{lumi}

\end{document}